\documentclass[aps,prl,showpacs,twocolumn,amsmath,amssymb]{revtex4-1}
\usepackage{epsfig}
\usepackage{amssymb}
\usepackage{multirow}
\usepackage{graphicx}

\begin{document} 

\title{Enhanced two dimensional electron gas charge densities at III-III/I-V oxide heterostructure interfaces} 

\author{Valentino R. Cooper}\email{coopervr@ornl.gov}
\affiliation{Materials Science and Technology Division, Oak Ridge National Laboratory, Oak Ridge, TN 37831-6056}

\begin{abstract} 
In this paper, density functional theory calculations are used to
explore the electronic and atomic reconstruction at interfaces between
III-III/I-V oxides.  In particular, at these
interfaces, two dimensional electron gases
(2DEGs) with twice the interfacial charge densities of the
prototypical LaTiO$_3$/SrTiO$_3$ heterostructure are observed.  Furthermore, a significant decrease in the band effective masses
of the conduction electrons is shown, suggesting that possible enhancements in electron mobilities may be achievable.  These findings
represent a framework for chemically modulating 2DEGs, thereby
providing a platform through which the underlying physics of
electron confinement can be explored with implications
for modern microelectronic devices.
\end{abstract}

\maketitle 

Emergent phenomena at \emph{AB}O$_3$ oxide interfaces, e.g.  two
dimensional electron gases (2DEGs),\cite{Ohtomo02p378} are paramount
to understanding critical behavior arising from electron confinement;
like metal-insulator transitions,\cite{Thiel06p1942} novel magnetic
effects \cite{Brinkman07p493} and
superconductivity.\cite{Klitzing80p494, Ando82p437} Recent theory and
experiments have sought to identify the origin of 2DEGs at insulating
oxide interfaces and surfaces in order to develop rules or concepts by
which charge carrier densities and mobilities can be tuned.  Growth
under low O$_2$ partial pressure has been shown to have (vacancy mediated) enhancements in conductivities that extend deep into bulk regions, making
these materials undesirable for examining interfacial physics.\cite{Siemons07p196802, Kalabukhov07p121404}
Conversely, annealing or growth under high O$_2$ pressure eliminates
oxygen vacancies resulting in consequential increases in resistivities; giving rise to intrinsic 2DEG behavior.\cite{Siemons07p196802, Kalabukhov07p121404} In oxide heterostructures such as LaAlO$_3$/SrTiO$_3$ and
LaGaO$_3$/SrTiO$_3$ this is understood in terms of polarization
discontinuities at interfaces, i.e. the \lq\lq{}\emph{the polar
  catastrophe}\rq\rq{} mechanism.\cite{Siemons07p196802, Ohtomo04p423,
  Nakagawa06p204, Pentcheva09p107602, Bristowe09p45425,
  Stengel09p241103} Physically, the divergence of the electrostatic
potential at atomically abrupt interfaces between insulating layered,
charge-ordered III-III  and a non-polar
II-IV (usually SrTiO$_3$) oxides is compensated for by charge
accumulation (in this case an extra $1/2$ electron per interface unit cell) at the
interface.  Interestingly enough, in these heterostructures this only occurs at TiO$_2$
centered interfaces and requires relatively thick layers of LaAlO$_3$
or LaGaO$_3$.\cite{Pentcheva09p107602, Perna10p152111}

A chemically intuitive, $\delta$-layer-doping, mechanism arises from the multivalent nature of transition metal
cations, like Ti.  For example, in LaTiO$_3$/SrTiO$_3$ superlattices, the
local environment of Ti cations next to a \lq\lq{}dopant\rq\rq{}, LaO layer,
splits the valence of Ti between two possible charge states (+4 for
SrTiO$_3$ and +3 for LaTiO$_3$).  Therefore, an equal
mixture of Ti valence states (3+ or 4+) can be thought to reside at
the interface giving an average valence of 3.5.\cite{Baraff77p237,
  Harrison78p4402, Chen10p2881} This has been confirmed through electron energy-loss spectroscopy (EELS)
measurements \cite{Jang11p886} in which the distribution of Ti$^{3+}$ cations away from the interface were found to be in
good agreement with theoretical and experimental carrier density
profiles.\cite{Siemons07p196802, Okamoto06p56802, Okamoto04p630,
  Santander-Syro11p189, Cantoni11p, Takizawa11p1} In a perfect system, the extra $1/2$
electron, relative to Ti in SrTiO$_3$, defines the intrinsic limit of 2DEG carrier densities.\cite{Kim10p201407} This electronic reconstruction is a hallmark of the observed
two-dimensional conductivity and is accompanied by polar
distortions \cite{Wang09p165130, Hamann06p195403,Okamoto06p56802}
(atomic displacements of the cations away from the interface) which
effectively screen the electrons near the interfaces.  Furthermore, the most
commonly studied 2DEGs derived from II-IV
(e.g. Sr$^{2+}$Ti$^{4+}$O$_3$) oxides have interfacial carrier densities which are
intrinsically limited to 0.5 e$^-$/interface unit cell.

In this paper, density functional theory (DFT) calculations are used
to investigate the charge rearrangement at interfaces between
I-V/III-III perovskites.  The goal is to apply the above insights to control the interfacial charge density by increasing the intrinsic limit of electrons at an oxide heterointerface.  The guiding principle is that the incorporation of a multivalent cation at an interface, where its desired valence states are +3 and +5, should
allow for an average valence of +4, thus increasing the limit of extra interfacial charge to 1 e$^-$/interface unit cell.  By examining heterostructures comprised of I-V/III-III oxides, it is confirmed that a total of 1 electron per
interface unit cell now populates the conduction bands (twice that of a corresponding LaTiO$_3$/SrTiO$_3$ superlattice).  Similar to
previous observations of 2DEGs, these electrons are primarily confined
to $t_{2g}$ orbitals on \emph{B} cations near LaO interfaces, decay
quickly into the bulk and are accompanied by large polar ionic distortions.  In addition, calculated decreases in electron band effective masses suggest that improved carrier mobilities may be achievable.

% Conceptually, heterostructures with multivalent
%cations with a larger valence difference should foster an increase in
%the intrinsic limit of the interfacial charge carrier density.  The question
%then remains as to how this density is redistributed throughout the
%heterostructure and whether or not this has any implications on the
%relative mobilities of the electron carriers.

DFT calculations using the local density approximation with
a Hubbard U (LDA+U) \cite{Anisimov91p943} and ultrasoft
pseudopotentials \cite{Vanderbilt90p7892} as implemented in the Quantum
Espresso simulation package \cite{Cococcioni05p35105,
  Giannozzi09p395502} were performed to study  1~La\emph{X}O$_3$/7~K\emph{X}O$_3$
superlattices, where \emph{X}=Ta and Nb.  All superlattice calculations employed an 80
Ry cutoff and an 8$\times$8$\times$1 k-point mesh.  Comparisons were
made with a prototypical 2DEG system, 1~LaTiO$_3$/7~SrTiO$_3$.  In all
calculations, the in-plane lattice constants were
constrained to the theoretical value of the majority component
(i.e. K\emph{X}O$_3$ or SrTiO$_3$) and the out-of-plane, $c$, lattice
vector was optimized within the P4mm space group with 1$\times$1
in-plane periodicity.  Simultaneously, all ionic coordinates were
relaxed until all Hellman-Feynman forces were less than 8 meV/\AA.
The computed bulk KNbO$_3$, KTaO$_3$, and SrTiO$_3$  cubic lattice
constants were 3.951~\AA, 3.945~\AA, and 3.855~\AA, respectively.
(Note: these values were obtained using standard LDA, i.e. without the
inclusion of a Hubbard U). These are in typical LDA agreement with
experimental values of 4.000~\AA, 3.988~\AA\ and 3.901~\AA,
respectively. For all heterostructure calculations, a Hubbard U=5 eV
for \emph{B}-cation $d$-states was found to be appropriate.  Similar U
values were  used in previous studies of LaTiO$_3$/SrTiO$_3$.\cite{Hamann06p195403, Okamoto06p56802} Band effective
masses were compute using quadratic fits of the partially occupied bands.

\begin{figure}[t]
\includegraphics[width=2.9in]{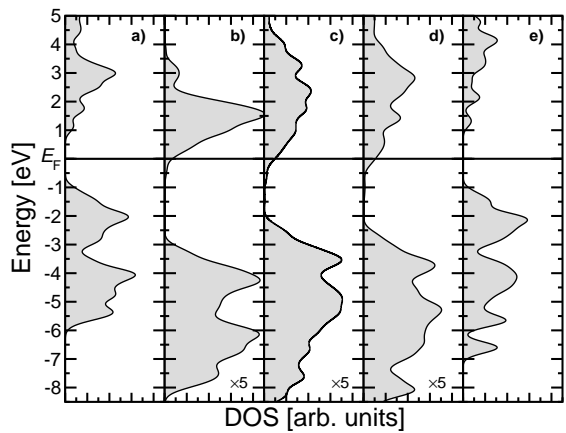}
\caption{\label{Fig1} DOS for (a) bulk SrTiO$_3$, (b)
  1~LaTiO$_3$/7~SrTiO$_3$, (c) 1~LaNbO$_3$/7~KNbO$_3$ (d)
  1~LaTaO$_3$/7~KTaO$_3$ and (e) bulk KTaO$_3$. All energies are
  relative to the Fermi level, $E_{\rm{F}}$.  (The scale for the
  heterostructures is 5$\times$ that of the bulk structures.)}
\end{figure}

\begin{figure}
\includegraphics[width=2.9in]{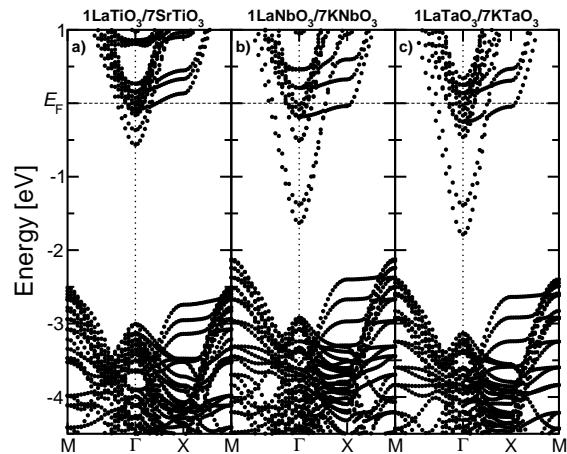}
\caption{\label{Fig2}Electronic band structure for (a)
  1~LaTiO$_3$/7~SrTiO$_3$, (b) 1~LaNbO$_3$/7~KNbO$_3$ and (c)
  1~LaTaO$_3$/7~KTaO$_3$ emphasizing the partially occupied states
  near the Fermi surface.}
\end{figure}

%Crucial to our understanding of the intrinsic charge carrier limit
%within the proposed heterostructures is an evaluation of the number of
%electrons available for conduction.  
Figure~\ref{Fig1} depicts the 
density of states (DOS) for bulk SrTiO$_3$, bulk KTaO$_3$,  1~LaTiO$_3$/7~SrTiO$_3$, 1~LaNbO$_3$ /7~KNbO$_3$ and
1~LaTaO$_3$/7~KTaO$_3$ superlattices.  Both SrTiO$_3$ and
KTaO$_3$, using LDA (i.e. no Hubbard U) have
relatively large band gaps of 1.7 and 1.8 eV, respectively.  Although
smaller than experiment, these are consistent with LDA\rq{}s
underestimation of oxide band gaps.  In agreement with previous
studies, 1~LaTiO$_3$/7~SrTiO$_3$ has occupied states
just below the Fermi level, $E_{\rm F}$, that sum to 1 electron (or rather 0.5 e$^-$/interface unit cell).  The electronic band structure
plot (Fig.~\ref{Fig2}a) indicates that they directly
contribute to transport (i.e. cross the Fermi level).
%In other words, these extra
%electrons are all conduction electrons.  
An analysis of the orbital projected DOS indicates that these states are derived mainly from Ti $t_{2g}$ states, with the two lowest energy, light electron, bands
coming almost entirely from $d_{\rm xy}$ orbitals on the interfacial Ti ions and the remaining occupied states being a mixture of $t_{2g}$ states on all of the Ti cations.  More importantly, in
1~LaNbO$_3$/7~KNbO$_3$ and 1~LaTaO$_3$/7~KTaO$_3$ these occupied
electronic states sum up to exactly 2 electrons (i.e. 1 e$^-$ /
interface unit cell).  An examination of the electronic band structure shows that they contribute to the Fermi surface (see
Fig.~\ref{Fig2}b and c). Orbital projected DOS indicate that they are
derived mainly from the \emph{B}-cation $d$-states (Nb/Ta) with
dominant electronic contributions arising from partially occupied
$d_{xy}$ orbitals of \emph{B}-cations at the LaO interface. In all
three heterostructures light electronic bands crossing $E_{\rm F}$ are
parabolic around $\Gamma$ and heavy bands extend along the $\Gamma$-X
direction.  This is a characteristic feature of 2DEGs and
is consistent with recent angle-resolved
photoemission spectroscopy (ARPES) results for 2DEGs at SrTiO$_3$
surfaces.\cite{Santander-Syro11p189, Meevasana11p114}

\begin{figure}
\includegraphics[width=2.9in]{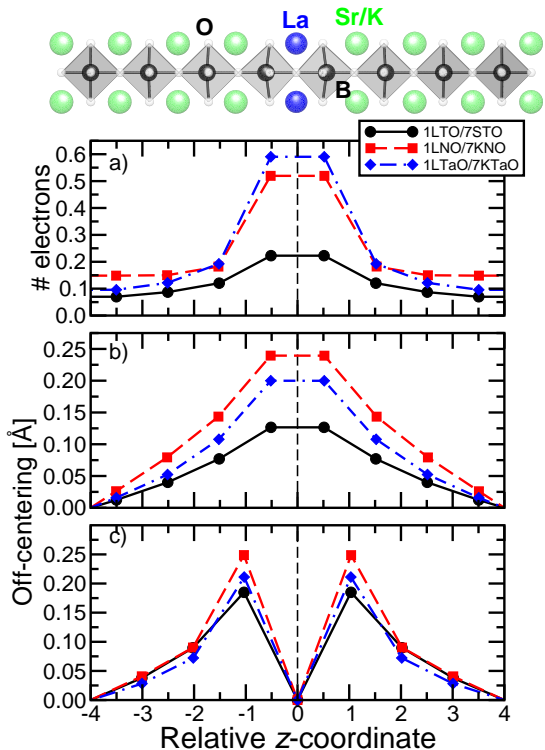}
\caption{\label{Fig3} (color online) [top] Representative
  2-dimensional projection of a relaxed superlattices.  (a) Charge distribution and (b) magnitude of
  \emph{B}- and (c) \emph{A}-cation off-centering as a function of relative
  $z$-coordinate for the superlattices studied.  All distances are relative to
  LaO planes.}
\end{figure}

%To analyze the spatial charge distribution we use atom projected DOS
%to compute the sum over states that contribute to the conduction
%electrons (i.e. states between $E_{\rm F}$ and $\sim$ -1.8 eV).
Figure~\ref{Fig3}a displays the spatial distribution of the conduction electrons (i.e. arising from states between $E_{\rm F}$ and $\sim$ -1.8 eV)
as a function of distance away from the LaO layer.  Similar to
previous theoretical and experimental results for 2DEGs at heterointerfaces  and
the SrTiO$_3$ surface we find a build-up of charge of roughly 0.22
electrons on interfacial Ti cations in the
1~LaTiO$_3$/7~SrTiO$_3$ superlattice.\cite{Siemons07p196802,
  Okamoto06p56802,  Okamoto04p630, Santander-Syro11p189, Cantoni11p,
  Takizawa11p1} (Note: the total atom projected DOS adds up to 0.95
e$^-$ and is scaled to unity in the plots.) This charge quickly decays
to 0.06 electrons in the center of the slab, indicating a decay length
of roughly 3-4 unit cells.  
%Theoretical studies employing a quasi-one
%dimensional, one-band model demonstrate that this decay length is
%related to the dielectric constant of the SrTiO$_3$
%layer.\cite{ Kancharla06p195427} 
On the other hand, we observe that the LaNbO$_3$/KNbO$_3$ and
LaTaO$_3$/KTaO$_3$ heterostructures have a build-up of 0.52 and 0.60
e$^-$s on the interfacial Nb and Ta ions, respectively.  (again the projected DOS only sums to ~1.8 e$^-$s and is uniformly
scaled to 2.0). Surprisingly, discernible
differences in the decay of charge away from the LaO
interface are seen. In LaNbO$_3$/KNbO$_3$, there is still roughly 0.15 e$^-$/unit cell area in the
bulk region, whereas the charge density in LaTaO$_3$/KTaO$_3$ drops
off much more sharply, falling to less than 0.1 e$^-$/unit cell
area. This deviation may be linked to differences in dielectric constants.  Theoretical calculations have
demonstrated that a larger dielectric constant can
result in a greater spread of electrons away from the
interface.\cite{Okamoto04p75101,Kancharla06p195427} KTaO$_3$ and SrTiO$_3$ have very
similar dielectric constants,\cite{Agrawal70p1120, Rupprecht64p748}
while KNbO$_3$ has a much stronger dependence of dielectric constant
on phase.\cite{Shirane54p672} In fact, KNbO$_3$ can have a dielectric
constants that is a few orders of magnitude greater than SrTiO$_3$ and
KTaO$_3$, making the observed behavior reasonable.  Regardless,
these results clearly indicate a significant enhancement in the concentration of electrons near the interface for
III-III/II-IV heterostructure relative to the III-III/I-V system, with thelower dielectric constant materials being more conducive to confining electrons to the interface.

In accordance with the observed interfacial electronic
reconstruction we find considerable polar distortions of the \emph{A}-
and \emph{B}-cations away from the LaO layer.  These off-center
displacements gradually return to zero in the center of the
SrTiO$_3$/K\emph{X}O$_3$ layers.  (See Fig.~\ref{Fig3} [top] for a
representative 2 dimensional projection of the atomic structure of a
1/7 heterostructure).  Similar to previous DFT results we find that the magnitude of SrTiO$_3$ off-centering is 0.18~\AA\ and 0.13~\AA\ for the \emph{A}- and \emph{B}-cations near the interface, respectively.\cite{Hamann06p195403, Okamoto06p56802} Figure~\ref{Fig3}b and c show that the
magnitudes of the \emph{B}- and \emph{A}-cation off-center displacements in the
La\emph{X}O$_3$/K\emph{X}O$_3$ structures are all appreciably larger
than those in LaTiO$_3$/SrTiO$_3$.   These larger polar distortions are consistent with the need
for larger interfacial polarizations to adequately screen the
interfacial charges.  Unexpectedly, the LaTaO$_3$/KTaO$_3$
superlattice while having the larger interfacial charge exhibits
smaller polar distortions than LaNbO$_3$/KNbO$_3$.  Although the
difference in the computed interfacial charge may be a consequence of
scaling of the charge densities, it may also be a direct indicator of
the magnitude of charge screening in these two materials arising from
the differences in their respective dielectric constants - with the
lower dielectric constant material being more strongly screened.  A
second explanation is that KNbO$_3$, unlike KTaO$_3$, has a polar
ground state which may be more favorable to inducing polar
distortions.  In which case, the polar nature of the KNbO$_3$
structure may be more advantageous as recent studies have proposed the
use of an electric field as a switch for controlling interface
conductivity.\cite{Wang09p165130, Caviglia08p624}

\begin{table}
\caption{\label{Tab1} Structural parameters and relative effective
  masses for the heterostructures studied.  $a_{\rm o}$, $c/a$ and
  $m_e/m$ denote the cubic lattice parameter, $c/a$ ratio and the
  relative effective masses of the two lowest energy partially
  occupied bands, respectively.  Note: the curvature of the electronic
  structure around $\Gamma$ is symmetric. i.e. the effective mass
  along the $\Gamma$-X and $\Gamma$-M directions are essentially equal
  and thus only one value is reported for each band. }
\begin{tabular}{lccccc}
\hline
\hline
\multirow{2}{*}{System} & $a_{\rm o}$ [\AA]    & $c/a$  & \multicolumn{2}{c}{$m_e/m$} \\
& & & 1 & 2 \\
\hline
1~LaTiO$_3$/7~SrTiO$_3$ & 3.885 & 8.117 & 0.49 & 0.59 \\
1~LaNbO$_3$/7~KNbO$_3$ & 3.951 & 8.085 & 0.35& 0.41  \\
1~LaTaO$_3$/7~KTaO$_3$ & 3.945 & 8.009 & 0.30 & 0.35 \\
 \hline
 \hline
 \end{tabular}
 \end{table}
 
Finally, 
%we examine the electron band effective masses, $m_e$, in order to
%give a preliminary assessment of the effects on carrier
%mobilities.  
Table~\ref{Tab1} lists the relative band effective masses,
$m_e/m$, of the two lowest energy partially occupied bands (see
Fig.~\ref{Fig2}).  
%For the light bands
%we find that in all systems the curvature of the electronic structure
%around $\Gamma$ is symmetric. In other words, $m_e/m$ along $\Gamma$-X
%and $\Gamma$-M are essentially equal.  Hence, we only report one value
%for each band.  
Remarkably, the computed $m_e/m$ values for SrTiO$_3$,
are in excellent agreement with ARPES measurements performed for
SrTiO$_3$ surface 2DEGs (0.5 - 0.6
$m_e/m$).\cite{Meevasana11p114} While the origin of the 2DEGs at
SrTiO$_3$ surfaces is almost certainly due to a different mechanism
than the $\delta$-doped structures, this result is suggestive of 
characteristic electronic structure features. It should be pointed out
that La doped SrTiO$_3$ has considerably higher effective mass ($m_e/m > 1$) and that the band effective masses predicted here neglects correlation effects. 
Moreover, the 1~La\emph{X}O$_3$/7~K\emph{X}O$_3$ structures were found
to have significant decreases in $m_e$ for these bands (which dominate
the electron density at the interfaces), implying possible increases in carrier mobilities in the III-III/I-V
superlattices. Of course, semiconductor physics tells us that
increases in charge densities (as observed in the
1~La\emph{X}O$_3$/7~K\emph{X}O$_3$ based heterostructures) are often
accompanied by decreases in carrier mobilities.  Typically increased
carrier densities arise from increases in dopant concentrations, which
act as scattering centers thereby reducing the mean free path of the
carriers.  In the above materials the LaO layers could be considered
dopants (i.e. $\delta$-doped layers).  Since the concentration of La
remains constant at the interface there may be little change in the
scattering lifetimes, $\tau$.  Hence, deviations in electron
mobilities, $\mu_e$, may be more affected by changes in $m_e$ (where
$\mu_e = e \tau/m_e$).  Furthermore, recent dynamical-mean-field calculations of oxide heterostructure 2DEGs suggest that correlation effects are least affected by dopant concentrations and that the band effective masses have the most deviations.\cite{Okamoto11p}

In summary, using first principles methods, it is demonstrated that superlattices
comprising I-V (K$^{1+}$[Nb/Ta]$^{5+}$O$_3$) and III-III
(La$^{3+}$[Nb/Ta]$^{3+}$O$_3$) perovskites have twice the interfacial
charge densities of III-III/II-IV(SrTiO$_3$/LaTiO$_3$) superlattices. Here,
the flexibility of multivalent cations, like Nb and Ta, leads to an intrinsic limit of 1 e$^-$ per interface unit cell, twice the limit of previously studied III-III/II-IV and I-V/III-III~\cite{Murray09p100102, Wang09p165130} superlattices.
Also, changes in
electron effective masses (with no change in dopant levels) imply that further enhancements in mobilities may be achievable.  
As such, this research highlights a viable path for enhancing the properties of 2DEGs.  In addition, deviations in polar distortions and charge
redistribution emphasize the need for a better understanding of the
relationship between dielectric constant and electronic and atomic reconstructions.  In agreement with other
theoretical models,\cite{Okamoto04p75101, Kancharla06p195427} these results suggest
that lower dielectric constant materials (e.g. KTaO$_3$) should
be more ideal for realizing truly two-dimensional electronic gases.
Alternatively, polar oxides like KNbO$_3$ may be more suitable for applications where it is necessary to switch the charge carrier concentrations.  Naturally, synthetic limitations related to effectively reducing ions like Ta$^{5+}$ to Ta$^{3+}$ may exist. A feasible route may be to substitute interfacial Ta
cations with a cation, like V, that is more easily reduced.  Ultimately, these results present a chemically intuitive framework (in the absence of factors such as O vacancies) through which intrinsic carrier concentrations, and perhaps even carrier mobilities, of oxide heterostructure 2DEGs can be tuned and
 may be useful in device engineering~\cite{Mannhart10p1607} and in controlling
quantum phenomena due to electron confinement.

V.R.C. would like to acknowledge helpful discussions with C. Bridges, C. Cantoni, H. N. Lee, S. Okamoto, D. Parker, W. Siemons  and D. Xiao. This work was supported by the Materials Sciences and  Engineering Division, Office of Basic Energy Sciences, U.S. Department of Energy. This research used resources of the  National Energy Research Scientific Computing Center, supported by the Office of Science, U.S. Department of Energy under Contract  No. DEAC02-05CH11231.

\end{document}